# Ultrafast single-nanowire multi-terahertz spectroscopy with sub-cycle temporal resolution

M. Eisele[1]*, T. L. Cocker[1]*, M. A. Huber[1], M. Plankl[1], L. Viti[2], D. Ercolani[2], L. Sorba[2], M. S. Vitiello[2] and R. Huber[1]†

[1] Department of Physics, University of Regensburg, 93040 Regensburg, Germany
[2] NEST, CNR – Istituto Nanoscienze and Scuola Normale Superiore, 56127 Pisa, Italy

**Phase-locked ultrashort pulses in the rich terahertz (THz) spectral range[1-18] have provided key insights into phenomena as diverse as quantum confinement[7], first-order phase transitions[8,12], high-temperature superconductivity[11], and carrier transport in nanomaterials[1,6,13-15]. Ultrabroadband electro-optic sampling of few-cycle field transients[1] can even reveal novel dynamics that occur faster than a single oscillation cycle of light[4,8,10]. However, conventional THz spectroscopy is intrinsically restricted to ensemble measurements by the diffraction limit. As a result, it measures dielectric functions averaged over the size, structure, orientation and density of nanoparticles, nanocrystals or nanodomains. Here, we extend ultrabroadband time-resolved THz spectroscopy (20 – 50 THz) to the sub-nanoparticle scale (10 nm) by combining sub-cycle, field-resolved detection (10 fs) with scattering-type near-field scanning optical microscopy (s-NSOM)[16-26]. We trace the time-dependent dielectric function at the surface of a single photoexcited InAs nanowire in all three spatial dimensions and reveal the ultrafast (<50 fs) formation of a local carrier depletion layer.**

* These authors contributed equally to this work.
† e-mail: rupert.huber@physik.uni-regensburg.de

Combining time-resolved terahertz (THz) spectroscopy with nanometer spatial resolution promises exciting possibilities for studying ultrafast dynamics in single nanoparticles. Few-THz (0.1 – 10 THz, far-infrared) to multi-THz (10 – 100 THz, mid-infrared) frequencies are home to many low-energy elementary excitations in condensed matter[1,2], including collective lattice, charge, and spin excitations. One valuable feature of THz spectroscopy is electro-optic sampling, which measures the oscillating electric field of light[1-4]. It provides the absolute phase and amplitude information of a broadband polarization response with time resolution faster than a single optical oscillation cycle[3,4,10]. However, while some nanoscale information can be inferred[1], the spatial resolution of far-field THz spectroscopy is intrinsically limited to the scale of the probing wavelength ($\lambda$ = 3 – 3000 µm).

Ultrafast THz spectroscopy beyond the diffraction limit has been a longstanding goal. It has been demonstrated that coupling THz pulses to sharp metallic tips encodes subwavelength spatial information onto the scattered fields[16-26], which have been detected either by intensity-resolving measurements in the multi-THz range[19-26] or electro-optic sampling in the few-THz window[16-18]. None of these studies, however, has probed photoinduced dynamics in single particles with nanometer lateral dimensions. Conversely, time-integrated intensity detection has been used to measure multi-THz pulses scattered from photoexcited graphene[26] with a temporal resolution of 200 fs, limited by the duration of the THz probe pulse. Following a different scheme, ultrafast charging dynamics in single nanoparticles have recently been measured electronically on the 1 nm scale using terahertz scanning tunnelling microscopy (THz-STM)[15], where local currents are induced by few-THz field transients.

Here, we use electro-optic sampling of multi-THz pulses to directly trace the scattered electric near field with a 10-fs gate pulse, revealing the dynamics of the dielectric function at a nanowire surface with 10-nm spatial resolution. Our experiment marks the first field-resolved pump-probe spectroscopy on the nanoscale and introduces sub-nanoparticle spatial resolution to sub-cycle multi-THz studies. We apply our new technique to InAs nanowires, a prototypical sample for conventional THz spectroscopy[1,6,13,14]. Nanowires based on III-V semiconductors have been shown to operate as efficient THz sources[13], active elements in one-dimensional field-effect transistors[27], or nanoscale infrared lasers[28,29]. Such nanodevices rely on a detailed knowledge of femtosecond carrier dynamics and surface charge distributions. Field-resolved THz nanoscopy of InAs nanowires allows us to directly resolve these ultrafast local effects for the first time and sets the stage for future sub-cycle near-field experiments in a wide range of nanosystems.

Our setup is based on state-of-the-art, ultrastable, Er:fibre laser technology[30], which is used to generate the pump, probe, and electro-optic gate pulses (see Methods) as summarized in Fig. 1a. We probe the transient dielectric response of an InAs nanowire with phase-stable, 2.5-cycle multi-THz pulses (Fig. 1b and c) following photoinjection of free carriers by near-infrared pump pulses. The THz pulses are focused onto the apex of a metallic atomic force microscope (AFM) tip, where they are strongly confined in the optical near field. Nanoscale information is retrieved from the scattered pulses when the evanescent field extending from the tip apex is modified by the sample. To isolate the near-field response from background scattering, the tip is operated in tapping mode and the scattered electric-field waveform $E_3(t)$ or the scattered intensity $I_3$ is measured at the third demodulation order of the tapping frequency[18-21,23-26]. We map out $E_3(t)$ by electro-optic sampling and detect $I_3$ using a time-integrating mercury cadmium telluride (MCT) photodiode, which can also be used to perform standard Fourier transform infrared spectroscopy (FTIR)[26].

The particular InAs nanowire investigated in the following is shown in the AFM topography image in Fig. 1d. Near-field intensity maps of the nanowire (Fig. 1e) were measured as a function of the relative arrival time between the near-infrared pump and THz probe at the sample, $t_{pp}$. Upon photoexcitation, scattering from the central axis of the nanowire surface is strongly enhanced ($t_{pp}$ = +50 fs). The increased scattering is short-lived: At $t_{pp}$ = +150 fs the nanowire has dimmed significantly in the near-field image. In Supplementary Discussion 1 we show differential images, which act as maps of the photoinduced carrier density. Line scans over a metal test sample are shown in Fig. 1f, demonstrating 10-nm edge resolution.

Pump-probe intensity scans ($I_3(t_{pp})$, Fig. 2a) were taken at specific positions on the nanowire, denoted in Fig. 1d. The dynamics were found to be highly dependent on tip position. In the centre of the nanowire (Position 1) the evolution of $I_3(t_{pp})$ is characterized by a large peak at $t_{pp}$ = +50 fs followed by a decay composed of two distinct time constants. The first decay occurs over approximately 100 fs, close to the time resolution of the pump-probe intensity measurements (~60 fs). In contrast, at the extremities of the nanowire the peak height is reduced (Positions 2 and 3). No pump-probe dynamics are present on the diamond substrate (Reference position).

To investigate the origin of the observed intensity dynamics we extend time-resolved multi-THz spectroscopy to the sub-nanoparticle scale. The initial 100-fs decay of $I_3(t_{pp})$ is of particular interest, as it is faster than typical carrier recombination or trapping times in semiconductors. Using electro-optic sampling, we show that it is possible to ascertain the *local* dielectric function over time scales shorter than an oscillation cycle of the probe pulse: An ultrashort gate pulse samples the instantaneous THz electric field and an entire effective waveform is recorded with sub-cycle time resolution[1-4] by scanning the THz-pulse delay ($t_{EOS}$) while keeping the pump-gate delay ($t_{pg}$) fixed (see Supplementary Discussion 2).

To perform such two-time measurements on a single nanowire, the complete near-field THz waveform must be recorded at the third harmonic of the tip tapping frequency. High sensitivity is a prerequisite, as this signal corresponds to only 50 photons per THz pulse. Using optimized electro-optic sampling in GaSe, we detect electric field transients from a 10-nm-wide area (Fig. 1f) on the surface of the nanowire with a noise floor of less than one coherent photon per pulse, as shown in Fig. 2b (black curves). We can even resolve changes to the transients triggered by sample photoexcitation. For $t_{pg}$ > 0 fs, a phase shift emerges in the latter half of the transient and the pulse rings for multiple oscillation cycles, indicating a pump-induced resonance in our frequency bandwidth. Electric-field changes caused by photoexcitation are plotted as red curves in Fig. 2b. The corresponding two-time electric-field plot (Fig. 2c) shows the full evolution of the scattered waveform as a function of $t_{pg}$, including a clear shift of the oscillating polarization response (Fig. 2c, dashed lines) relative to the driving THz field, with increasing $t_{pg}$.

Field-resolved detection of the scattered THz waveform provides access to both the absolute phase and amplitude in the frequency domain (Fig. 3). The resonance manifests itself as a dip in the amplitude spectrum of $\tilde{E}_3$ at frequency $f_0$ accompanied by a shift in its phase. Interestingly, $f_0$ redshifts for $t_{pg} > +50$ fs (Fig. 3a-d). We simulate the scattered spectra using the point-dipole model[16] and a nanowire dielectric function defined by the Drude model. Using only the carrier density ($N_c$) and scattering time ($\tau$) as free parameters, we reproduce the time evolution of both the amplitude (Fig. 3a and b) and absolute phase (Fig. 3c and d) of $\tilde{E}_3$. Within this framework $f_0 \approx \omega_p/2\pi$, where $\omega_p \propto \sqrt{N_c}$ is the plasma frequency of free carriers in the nanowire. Thus, $f_0$ directly tracks the local carrier density following photoexcitation. The resonance position is plotted as a function of $t_{pg}$ in Fig. 3e, while the carrier densities extracted from the point-dipole model are shown in Fig. 3f. The two distinct time constants observed in the intensity dynamics also appear in the dynamics of $f_0$ and $N_c$. At early times, both quantities drop rapidly, with time constants significantly faster than the resolution-limited decay observed in the intensity measurements. In fact, the time constant of the decay of $f_0$ extracted directly from the electro-optic spectra reaches 40±10 fs. The same behavior can alternatively be characterized by FTIR with reduced time resolution (>60 fs, see Supplementary Discussion 5).

The near-field confinement that enables 10-nm lateral resolution in our measurements also applies to the field extending vertically from the tip apex into the sample. Interestingly, the tapping amplitude can be used to tune the effective decay length in free space of the scattered evanescent near field sampled at the third demodulation order, as shown in Fig. 4a. Equivalently, the tapping amplitude controls the probing depth into the sample when the tip is approached[19]. We exploit this dependence to develop a new technique — femtosecond tomography — that allows us to resolve depth-dependent dynamics and identify the source of the 40-fs decay. Figure 4b shows the effect of the tapping amplitude on the resonance frequency 300 fs after photoexcitation. For tapping amplitudes below 70 nm, $f_0$ exhibits a distinct redshift, implying a reduced carrier density at shallower probing depths (depletion layer). Surprisingly, the pump-probe intensities at $t_{pp}$ = +50 fs coincide for all tapping amplitudes (Fig. 4c), indicating that the carrier distribution is homogeneous immediately following photoexcitation. For a comprehensive picture, we directly trace the evolution of $f_0$ for low (50 nm) and high (130 nm) tapping amplitudes (Fig. 4d). The data confirm that the fast initial decay of $f_0$ corresponds to the formation of the depletion layer: For sufficiently small tapping amplitudes (50 nm) a built-in surface field[31,32] accelerates photoinduced carriers out of our probing volume (estimated thickness: ~10 nm). After 200 fs the depletion layer is fully formed, leading to a ~10 % difference in average carrier density between the probing volumes for low and high tapping amplitudes. The carrier densities subsequently decay on approximately the same time scale for low and high tapping amplitudes (2 ps) due to carrier trapping into defect states. Measurements of surface depletion layer formation at the end of the nanowire (Position 3, Fig. 1d) are shown in Supplementary Discussion 7, as is an estimate of the surface field.

In summary, we have combined ultrabroadband field-resolved detection of multi-THz pulses with s-NSOM to enable single-nanoparticle THz spectroscopy with simultaneous 10-nm spatial resolution and sub-cycle, 10-fs temporal resolution. Our novel system has been applied to probe the photoinduced dynamics in single InAs nanowires, where we have also developed femtosecond tomography to observe ultrafast depletion layer formation. Multi-THz spectroscopy of single nanoparticles with both sub-cycle time resolution and three-dimensional local field sensitivity opens up a new world for THz spectroscopy far below the diffraction limit ($\lambda^3/10^9$). Nanoscale experiments completely free of effective medium theories can now be envisioned for virtually any physical, chemical and biological process appropriate for time-resolved spectroscopy in the multi-THz range.

## Materials and Methods

### InAs nanowires

Indium arsenide (InAs, $E_g \approx 0.35$ eV) nanowires are employed as a model system for our novel sub-cycle nanoscope. InAs features a high electron mobility (6000 cm$^2$/Vs) as well as a potentially long electron mean free path. InAs nanowires hold promise for future integration with low-cost silicon technology[27]. Self assembled nanowires were grown bottom-up on InAs (111)B substrates by chemical beam epitaxy (CBE) in a Riber Compact-21 system using gold particles as a growth catalyst. Tri-methyl-indium (TMIn) and tertiary-butyl-arsine (TBAs) were used as metal–organic chemical precursors, while di-tertiary-butyl-selenide (DTBSe) served as a selenium source for n-type doping. A 0.5-nm-thick Au film was first deposited on the InAs wafer by thermal evaporation. The wafer was then transferred to the CBE system and annealed at 520 °C under TBAs flow in order to remove the surface oxide and generate the Au nanoparticles by thermal dewetting. The InAs segment was grown for 90 min at a temperature of (430±10) °C, with metal-organic line pressures of 0.3 and 1.0 Torr for TMIn and TBAs, respectively. The DTBSe line pressure was fixed at 0.1 Torr to achieve n-type doping with a density of $N_c \approx 10^{17}$ cm$^{-3}$. The nanowires were then mechanically transferred to a diamond substrate.

### Ultrafast near-field nanoscopy setup

The laser system used for ultrafast near-field microscopy is based on four separate erbium-doped fibre (Er:fibre) amplifiers seeded by a common oscillator (Fig. 1a), resulting in four mutually coherent pulse trains operating at repetition rates of 20 - 40 MHz with pulse energies of 9 - 15 nJ (Toptica). The pulses from each amplifier are spectrally shaped by nonlinear fibres: amplifier 1, which is used to pump the sample, operates at 40 MHz and produces 22-fs-long (FWHM), 5 nJ near-infrared pulses centered at 192 THz after spectral shaping. The multi-terahertz (multi-THz) probe pulses are generated by critically phase-matched, non-collinear difference frequency generation of the pulses from amplifier 2 (centre frequency 153 THz, pulse energy 1.5 nJ, pulse length 30 fs) and amplifier 3 (centre frequency 192 THz, pulse energy 5 nJ, pulse length 30 fs) in a 1-mm-thick gallium selenide (GaSe) crystal. By adjusting the phase-matching condition in GaSe we can spectrally tune the multi-THz pulses to cover the frequency range from 15 THz to 60 THz with a pulse length of 60 fs and a pulse energy of 30 pJ.

Sub-cycle detection of THz waveforms is attained via electro-optic sampling. The THz pulses are focused into a GaSe crystal (thickness: 180 µm) where the instantaneous THz-field-induced birefringence is encoded in the polarization rotation of gate pulses supplied by amplifier 4 (centre frequency 230 THz, pulse energy 0.8 nJ, pulse length 10 fs). By scanning the THz-gate delay we directly map the oscillating THz electric field in the time domain.

Subwavelength spatial resolution is achieved by focusing the multi-THz pulses onto the sharp metallic tip (Pt/Ir coating, grounded) of a scattering-type near-field scanning optical microscope (s-NSOM, Neaspec) operated in tapping mode (frequency 270 kHz). The tip acts like an ultrabroadband antenna that strongly confines electromagnetic radiation in the near field of the tip apex. The tip-sample polarization response leads to a scattered electromagnetic pulse carrying information about the dielectric function of the sample with 10-nm spatial resolution. The scattered field is detected electro-optically providing absolute phase and amplitude information. Alternatively, the scattered intensity is recorded by a mercury cadmium telluride photodiode that can be used for Fourier transform infrared spectroscopy (FTIR). FTIR provides the phase relative to the phase of a reference pulse and the amplitude convoluted with the amplitude of the reference pulse. To suppress background radiation from the tip shaft and from the sample, the signal demodulated at the 3rd harmonic of the tapping frequency is detected with the help of a lock-in amplifier. The pump pulses are focused to a 10-µm-diameter spot on the sample below the tip apex using the same parabolic mirror that focuses the probe pulses.

**References**

1. Ulbricht, R., Hendry, E., Shan, J., Heinz, T. F. & Bonn, M. Carrier dynamics in semiconductors studied with time-resolved terahertz spectroscopy. *Rev. Mod. Phys*. **83**, 543-586 (2011).

2. Kampfrath, T., Tanaka, K. & Nelson, K. A. Resonant and nonresonant control over matter and light by intense terahertz transients. *Nature Photon.* **7**, 680-690 (2013).

3. Kindt, J. T. & Schmuttenmaer, C. A. Theory for determination of the low-frequency time-dependent response function in liquids using time-resolved terahertz pulse spectroscopy. *J. Chem. Phys.* **110**, 8589-8596 (1999).

4. Huber, R. *et al*. How many-particle interactions develop after ultrafast excitation of an electron-hole plasma. *Nature* **414**, 286-289 (2001).

5. Kaindl, R. A., Carnahan, M. A., Hägele, D., Lövenich, R. & Chemla, D. S. Ultrafast terahertz probes of transient conducting and insulating phases in an electron-hole gas. *Nature* **423**, 734-738 (2003).

6. Baxter, J. B. & Schmuttenmaer, C. A. Conductivity of ZnO Nanowires, Nanoparticles, and Thin Films Using Time-Resolved Terahertz Spectroscopy. *J. Phys. Chem. B* **110**, 25229-25239 (2006).

7. Wang, F. *et al*. Exciton polarizability in semiconductor nanocrystals. *Nature Mater.* **5**, 861-864 (2006).

8. Kübler, C. *et al*. Coherent Structural Dynamics and Electronic Correlations during an Ultrafast Insulator-to-Metal Phase Transition in $VO_2$. *Phys. Rev. Lett.* **99**, 116401 (2007).

9. Gaal, P. *et al*. Internal motions of a quasiparticle governing its ultrafast nonlinear response. *Nature* **450**, 1210-1213 (2007).

10. Günter, G. *et al*. Sub-cycle switch-on of ultrastrong light-matter interaction. *Nature* **458**, 178-181 (2009).

11. Pashkin, A. *et al*. Femtosecond Response of Quasiparticles and Phonons in Superconducting $YBa_2Cu_3O_{7-\delta}$ Studied by Wideband Terahertz Spectroscopy. *Phys. Rev. Lett.* **105**, 067001 (2010).

12. Liu, M. K. *et al*. Photoinduced Phase Transitions by Time-Resolved Far-Infrared Spectroscopy in $V_2O_3$. *Phys. Rev. Lett.* **107**, 066403 (2011).

13. Seletskiy, D. V. *et al*. Efficient terahertz emission from InAs nanowires. *Phys. Rev. B* **84**, 115421 (2011).

14. Joyce, H. J. *et al*. Electronic properties of GaAs, InAs and InP nanowires studied by terahertz spectroscopy. *Nanotechnology* **24**, 214006 (2013).

15. Cocker, T. L. *et al*. An ultrafast terahertz scanning tunnelling microscope. *Nature Photon*. **7**, 620-625 (2013).

16. Chen, H.-T., Kersting, R. & Cho, G. C. Terahertz imaging with nanometer resolution. *Appl. Phys. Lett.* **83**, 3009-3011 (2003).

17. Zhan, H. *et al*. The metal-insulator transition in $VO_2$ studied using terahertz apertureless near-field microscopy. *Appl. Phys. Lett.* **91**, 162110 (2007).

18. Moon, K. *et al*. Quantitative coherent scattering spectra in apertureless terahertz pulse near-field microscopes. *Appl. Phys. Lett.* **101**, 011109 (2012).

19. Krutokhvostov, R. *et al*. Enhanced resolution in subsurface near-field optical microscopy. *Opt. Express* **20**, 593-600 (2011).

20. Qazilbash, M. M. *et al*. Mott Transition in $VO_2$ Revealed by Infrared Spectroscopy and Nano-Imaging. *Science* **318**, 1750-1753 (2007).

21. Jones, A. C. *et al*. Mid-IR Plasmonics: Near-Field Imaging of Coherent Plasmon Modes of Silver Nanowires. *Nano Lett.* **9**, 2553-2558 (2009).

22. Diyar, S. *et al*., Adiabatic Nanofocusing Scattering-Type Optical Nanoscopy of Individual Gold Nanoparticles. *Nano Lett.* **11**, 1609-1613 (2011).

23. Chen, J. *et al*. Optical nano-imaging of gate-tunable graphene plasmons. *Nature* **487**, 77-81 (2012).

24. Fei, Z. *et al*. Gate-tuning of graphene plasmons revealed by infrared nano-imaging. *Nature* **487**, 82-85 (2012).

25. Jacob, R. *et al*. Intersublevel Spectroscopy on Single InAs-Quantum Dots by Terahertz Near-Field Microscopy. *Nano Lett.* **12**, 4336-4340 (2012).

26. Wagner, M. *et al*. Ultrafast and Nanoscale Plasmonic Phenomena in Exfoliated Graphene Revealed by Infrared Pump-Probe Nanoscopy. *Nano Lett.* **14**, 894-900 (2014).

27. Vitiello, M. S. *et al*. Room-Temperature Terahertz Detectors Based on Semiconductor Nanowire Field-Effect Transistors. *Nano Lett.* **12**, 96-101 (2012).

28. Saxena, D. *et al*. Optically pumped room-temperature GaAs nanowire lasers. *Nature Photon.* **7**, 963-968 (2013).

29. Mayer, B. *et al*. Lasing from individual GaAs-AlGaAs core-shell nanowires up to room temperature. *Nature Commun.* **4**, 2931 (2013).

30. Krauss, G. *et al*. Synthesis of a single cycle of light with compact erbium-doped fibre technology. *Nature Photon.* **4**, 33-36 (2010).

31. Piper, L. F. J., Veal, T. D., Lowe, M. J. & McConville, C. F. Electron depletion at InAs free surfaces: Doping-induced acceptorlike gap states. *Phys. Rev. B* **73**, 195321 (2006).

32. Dekorsy, T., Pfeifer, T., Kütt, W. & Kurz, H. Subpicosecond carrier transport in GaAs surface-space-charge fields. *Phys. Rev. B.* **47**, 3842-3849 (1993).

**Acknowledgements**

We thank M. Furthmeier for technical assistance and J. Lupton, P. Klemm, D. Bougeard, B. Surrer, R. Hillenbrand and F. Keilmann for valuable discussions. This work was supported by the European Research Council through ERC grant 305003 (QUANTUMsubCYCLE) and the Graduate Research College (GRK 1570). T.L.C acknowledges the support of the Alexander von Humboldt Foundation.

**Author contributions**

M.E., T.L.C., and R.H. conceived the study and build the experimental setup. M.E., T.L.C., M.A.H., M.P. and R.H. carried out the experiment and analyzed the data. M.A.H., T.L.C. and M.E. performed simulations. L.V., D.E., L.S., and M.S.V. grew the InAs nanowires. T.L.C., M.E., M.A.H., and R.H. wrote the manuscript. All authors contributed to the discussions.

**Additional information**

Correspondence and requests for materials should be addressed to R.H.

**Competing financial interests**

The authors declare no competing financial interests.

**Figure captions**

**Figure 1 | Setup for single-nanowire terahertz spectroscopy. a**, Schematic of the experimental setup. A femtosecond erbium-doped fibre (Er:fibre) oscillator seeds four separate Er:fibre amplifiers that are used to produce the near-infrared (1.56 μm) pump pulses (Amp 1), near-infrared (1.3 μm) electro-optic gate pulses (Amp 4), and phase-stable multi-THz probe pulses (Amp 2 and Amp 3) via difference frequency generation (DFG). The THz transients are focused onto the AFM tip of a scattering-type near-field scanning optical microscope (s-NSOM) and the scattered electric near-field is detected by electro-optic sampling (EOS) with sub-cycle temporal resolution. Alternatively, a mercury cadmium telluride (MCT) photodiode records the time-integrated scattered intensity and enables Fourier transform infrared spectroscopy (FTIR, >60 fs time resolution). **b**, Electric field of THz reference pulse focused onto the s-NSOM tip, detected by electro-optic sampling. **c**, Amplitude (black curve) and absolute phase (red curve) of THz reference waveform. **d**, Topography of the indium arsenide (InAs) nanowire studied in our experiments, measured by atomic force microscopy. Substrate: diamond **e**, Ultrafast THz nano-movie of nanowire photoexcitation. Free carriers are photoinjected into the InAs nanowire by near-infrared pump pulses and time-resolved near-field THz intensity images are measured as a function of pump-probe delay time $t_{pp}$. The pump fluence was 1.0 mJ/cm$^2$ and the tapping amplitude was 130 nm. **f**, Scattered near-field intensity over a rough metal (silver) test sample recorded directly by an MCT photodiode (black curve, top) and extracted by spectrally integrating and squaring electric-field waveforms measured at each *x*-position by EOS (red curve, top). Bottom: corresponding topography. Grey dashed line: guide to the eye. Black arrows: 10-nm-wide edges.

**Figure 2 | Dynamics of the oscillating electric near field. a**, Pump-probe intensity scans measured at various sites on the nanowire (Position 1 – 3, see Fig. 1d) and on the diamond substrate (Reference position). The rise time of the sharp peak is ~60 fs. **b**, THz electric-field waveforms measured by electro-optic sampling in the centre of the nanowire (Position 1) at different pump-gate delay times (black curves). All waveforms were recorded by scanning the THz-gate delay ($t_{EOS}$) with fixed pump-gate delay ($t_{pg}$). The scattered waveform at negative delay time ($t_{pg}$ = -5 ps, top) is strongly altered by sample excitation, as can be seen in the differential waveforms $E_3(t_{pg})$ - $E_3$(-5 ps) (red curves). **c**, Two-time map of the scattered electric near field at the centre of the nanowire. Black dashed lines are guides to the eye. For all data the pump fluence was 1.1 mJ/cm$^2$ and the tapping amplitude was 130 nm.

**Figure 3 | Sub-cycle spectral dynamics. a**, Amplitude spectra of THz electric-field waveforms measured in the centre of the nanowire (Position 1, Fig. 1d) as a function of pump-gate delay ($t_{pg}$). **b**, Theoretical scattered amplitudes simulated with the point-dipole model and a dielectric function given by the Drude model. See Supplementary Discussion 3 for simulation details. **c**, Absolute phases of scattered waveforms measured by EOS. A phase shift of ∼2.5 rad is observed at the resonance. **d**, Theoretical phases extracted from the point-dipole model. Blue points in **a-d** mark the dip frequencies $f_0$ in **a**, and blue curves serve as guides to the eye. **e**, Resonance frequency $f_0$ extracted from **a** as a function of $t_{pg}$, plotted on a semi-logarithmic scale. A crossover between fast (40 fs) and slow (∼4 ps) dynamics occurs at $t_{pg}$ = +200 fs, pointing towards two distinct physical mechanisms in the temporal evolution of the carrier density. The error bars are estimated based on the spectral resolution and the widths of the minima. **f**, Carrier density $N_c$ extracted from the simulations in **b** and **d** plotted on a semi-logarithmic scale. The corresponding scattering rates are shown in Supplementary Discussion 4. As in $f_0(t_{pg})$, a crossover is observed between two different time constants (20 fs and 2 ps) at $t_{pg}$ = +200 fs. Red dashed lines in **e** and **f** are exponential decays included as guides for the eye. Pump fluence: 1.1 mJ/cm$^2$; tapping amplitude: 130 nm.

**Figure 4 | Femtosecond tomography. a**, Decay distance (1/e) of scattered intensity as a function of tapping amplitude extracted from retraction scans over a gold reference sample. The grey dashed line is a linear fit to the data. We estimate the probing depth into the nanowire to be the free-space decay length divided by the index of refraction of InAs. The error bars correspond to 95% confidence intervals. **b**, Resonance frequency $f_0$ as a function of tapping amplitude (TA) at $t_{pp}$ = +300 fs, reflecting changes to the carrier density as a function of probing depth. Grey dashed lines are guides to the eye. **c**, Pump-probe dynamics of the scattered intensity showing a strong dependence on tapping amplitude. Scans are normalized to the unpumped baseline at $t_{pp}$ = -5 ps to account for changes to scattering efficiency with tapping amplitude. **d**, Resonance frequencies extracted from measured FTIR spectra as a function of time after photoexcitation for TA = 130 nm (black points) and TA = 50 nm (red points) plotted on a semi-logarithmic graph. The resonance frequency is reduced for the low tapping amplitude when a depletion layer is present. Grey dashed lines are exponential guides to the eye. The error bars in **b** and **d** are estimated as in Fig. 3e. Pump fluence: 0.75 mJ/cm$^2$. Similar $f_0$ dynamics are shown in Supplementary Discussion 6 for pump fluences of 0.5 mJ/cm$^2$ and 1.0 mJ/cm$^2$.

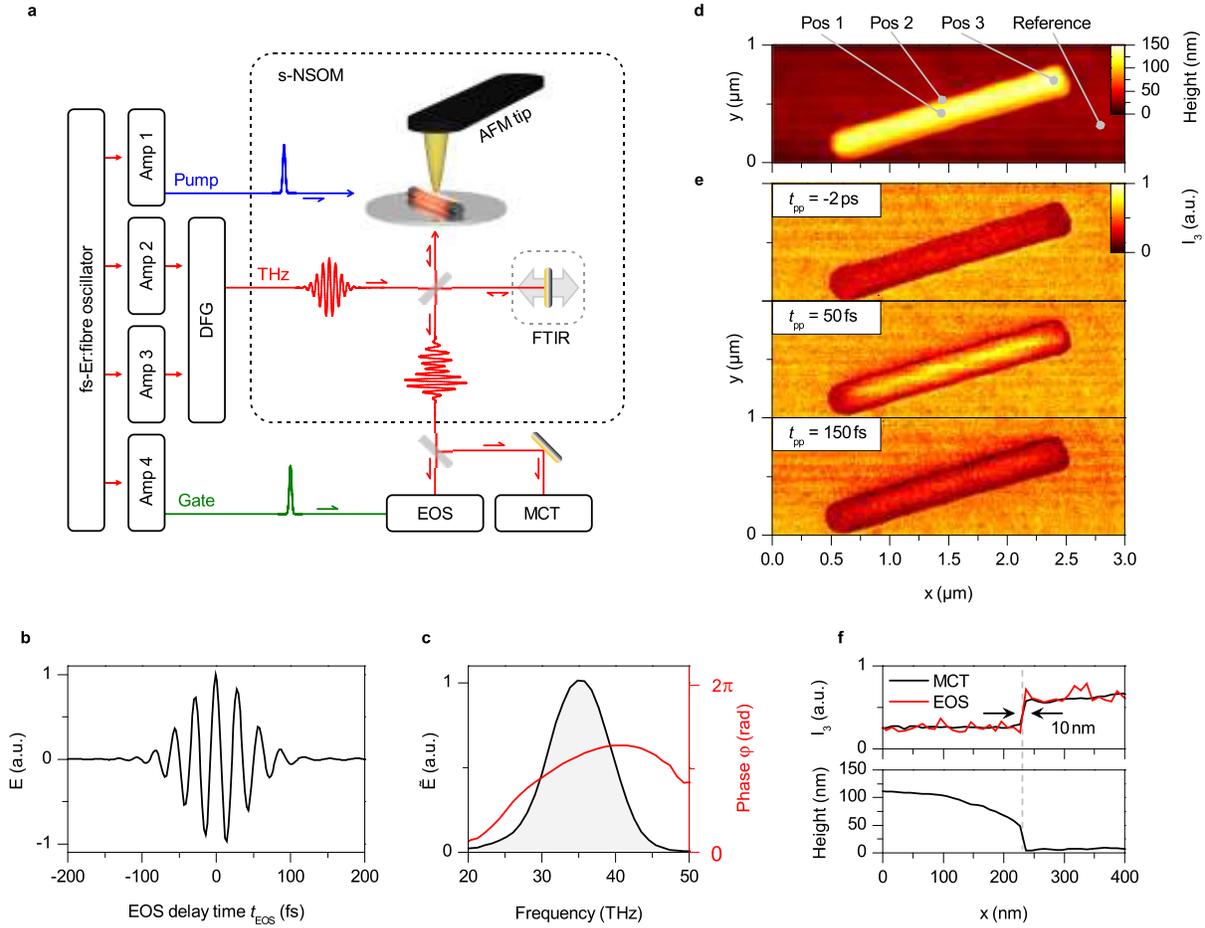

**Figure 1 | Setup for single-nanowire terahertz spectroscopy. a**, Schematic of the experimental setup. A femtosecond erbium-doped fibre (Er:fibre) oscillator seeds four separate Er:fibre amplifiers that are used to produce the near-infrared (1.56 µm) pump pulses (Amp 1), near-infrared (1.3 µm) electro-optic gate pulses (Amp 4), and phase-stable multi-THz probe pulses (Amp 2 and Amp 3) via difference frequency generation (DFG). The THz transients are focused onto the AFM tip of a scattering-type near-field scanning optical microscope (s-NSOM) and the scattered electric near-field is detected by electro-optic sampling (EOS) with sub-cycle temporal resolution. Alternatively, a mercury cadmium telluride (MCT) photodiode records the time-integrated scattered intensity and enables Fourier transform infrared spectroscopy (FTIR, >60 fs time resolution). **b**, Electric field of THz reference pulse focused onto the s-NSOM tip, detected by electro-optic sampling. **c**, Amplitude (black curve) and absolute phase (red curve) of THz reference waveform. **d**, Topography of the indium arsenide (InAs) nanowire studied in our experiments, measured by atomic force microscopy. Substrate: diamond **e**, Ultrafast THz nano-movie of nanowire photoexcitation. Free carriers are photoinjected into the InAs nanowire by near-infrared pump pulses and time-resolved near-field THz intensity images are measured as a function of pump-probe delay time $t_{pp}$. The pump fluence was 1.0 mJ/cm$^2$

and the tapping amplitude was 130 nm. **f**, Scattered near-field intensity over a rough metal (silver) test sample recorded directly by an MCT photodiode (black curve, top) and extracted by spectrally integrating and squaring electric-field waveforms measured at each *x*-position by EOS (red curve, top). Bottom: corresponding topography. Grey dashed line: guide to the eye. Black arrows: 10-nm-wide edges.

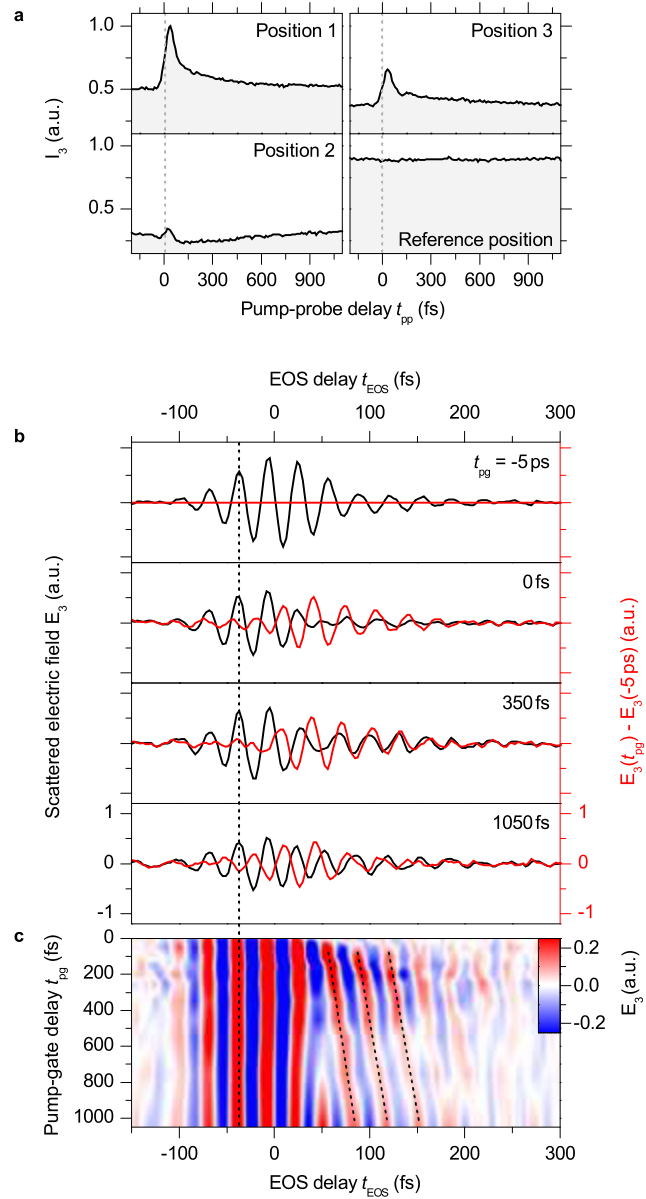

**Figure 2 | Dynamics of the oscillating electric near field. a**, Pump-probe intensity scans measured at various sites on the nanowire (Position 1 – 3, see Fig. 1d) and on the diamond substrate (Reference position). The rise time of the sharp peak is ∼60 fs. **b**, THz electric-field waveforms measured by electro-optic sampling in the centre of the nanowire (Position 1) at different pump-gate delay times (black curves). All waveforms were recorded by scanning the THz-gate delay ($t_{EOS}$) with fixed pump-gate delay ($t_{pg}$). The scattered waveform at negative delay time ($t_{pg}$ = -5 ps, top) is strongly altered by sample excitation, as can be seen in the differential waveforms $E_3(t_{pg})$ - $E_3$(-5 ps) (red curves). **c**, Two-time map of the scattered electric near field at the centre of the nanowire. Black dashed lines are guides to the eye. For all data the pump fluence was 1.1 mJ/cm² and the tapping amplitude was 130 nm.

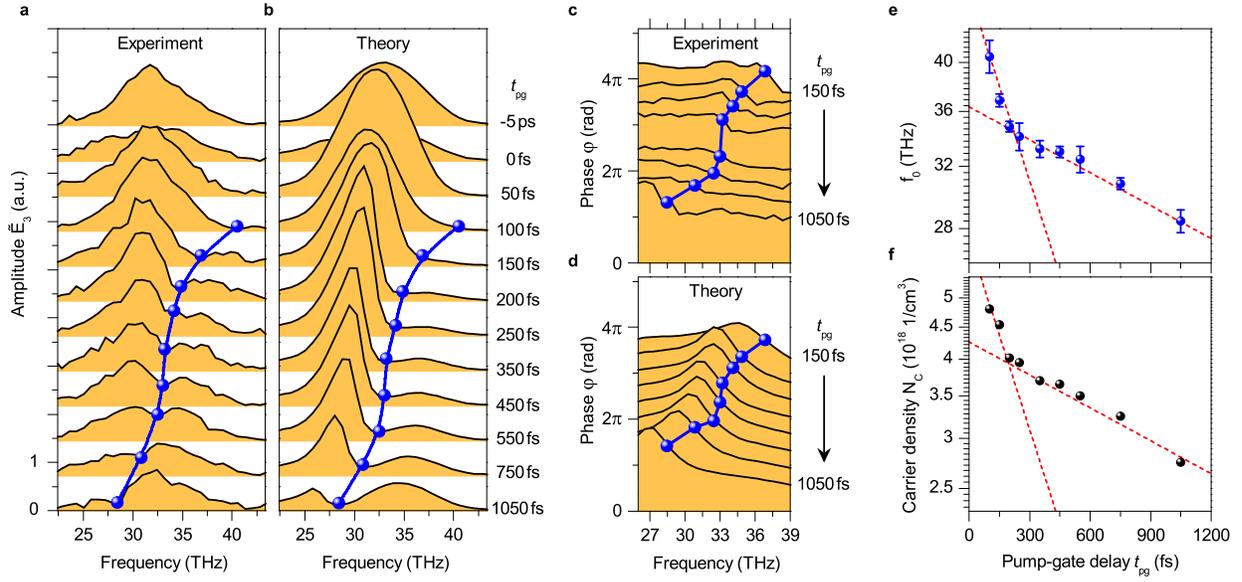

**Figure 3 | Sub-cycle spectral dynamics. a**, Amplitude spectra of THz electric-field waveforms measured in the centre of the nanowire (Position 1, Fig. 1d) as a function of pump-gate delay ($t_{pg}$). **b**, Theoretical scattered amplitudes simulated with the point-dipole model and a dielectric function given by the Drude model. See Supplementary Discussion 3 for simulation details. **c**, Absolute phases of scattered waveforms measured by EOS. A phase shift of ~2.5 rad is observed at the resonance. **d**, Theoretical phases extracted from the point-dipole model. Blue points in **a-d** mark the dip frequencies $f_0$ in **a**, and blue curves serve as guides to the eye. **e**, Resonance frequency $f_0$ extracted from **a** as a function of $t_{pg}$, plotted on a semi-logarithmic scale. A crossover between fast (40 fs) and slow (~4 ps) dynamics occurs at $t_{pg}$ = +200 fs, pointing towards two distinct physical mechanisms in the temporal evolution of the carrier density. The error bars are estimated based on the spectral resolution and the widths of the minima. **f**, Carrier density $N_c$ extracted from the simulations in **b** and **d** plotted on a semi-logarithmic scale. The corresponding scattering rates are shown in Supplementary Discussion 4. As in $f_0(t_{pg})$, a crossover is observed between two different time constants (20 fs and 2 ps) at $t_{pg}$ = +200 fs. Red dashed lines in **e** and **f** are exponential decays included as guides for the eye. Pump fluence: 1.1 mJ/cm$^2$; tapping amplitude: 130 nm.

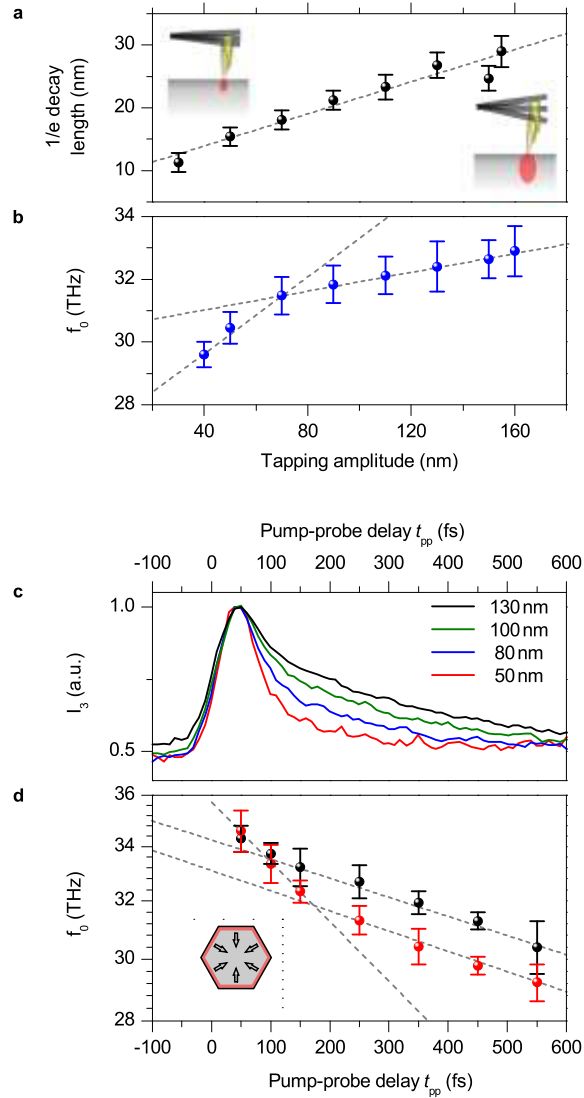

**Figure 4 | Femtosecond tomography. a**, Decay distance (1/e) of scattered intensity as a function of tapping amplitude extracted from retraction scans over a gold reference sample. The grey dashed line is a linear fit to the data. We estimate the probing depth into the nanowire to be the free-space decay length divided by the index of refraction of InAs. The error bars correspond to 95% confidence intervals. **b**, Resonance frequency $f_0$ as a function of tapping amplitude (TA) at $t_{pp}$ = +300 fs, reflecting changes to the carrier density as a function of probing depth. Grey dashed lines are guides to the eye. **c**, Pump-probe dynamics of the scattered intensity showing a strong dependence on tapping amplitude. Scans are normalized to the unpumped baseline at $t_{pp}$ = -5 ps to account for changes to scattering efficiency with tapping amplitude. **d**, Resonance frequencies extracted from measured FTIR spectra as a function of time after photoexcitation for TA = 130 nm (black points) and TA = 50 nm (red points) plotted on a semi-logarithmic graph. The resonance frequency is reduced for the low tapping amplitude when a depletion layer is present. Grey dashed lines are exponential guides to

the eye. The error bars in **b** and **d** are estimated as in Fig. 3e. Pump fluence: 0.75 mJ/cm². Similar $f_0$ dynamics are shown in Supplementary Discussion 6 for pump fluences of 0.5 mJ/cm² and 1.0 mJ/cm².

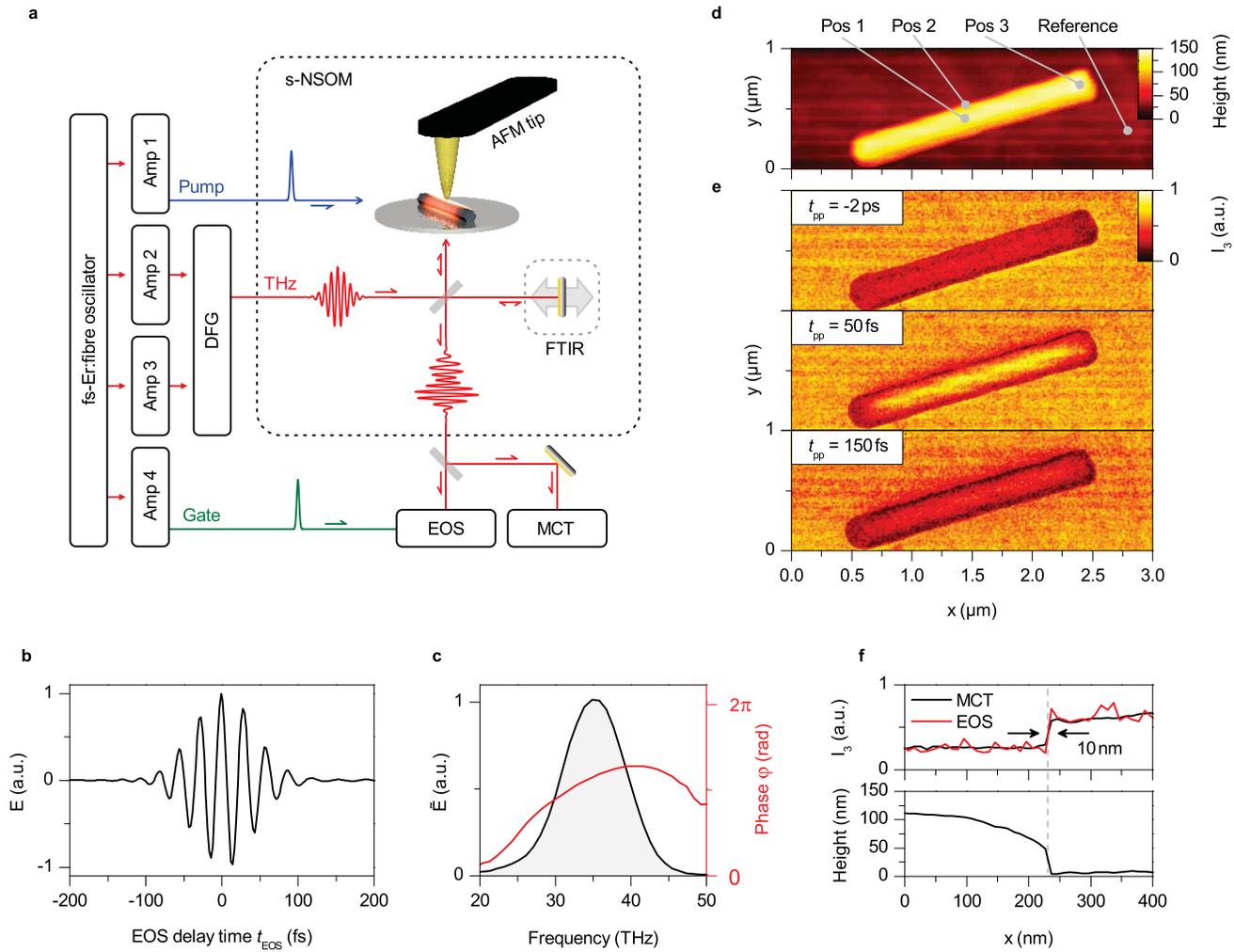

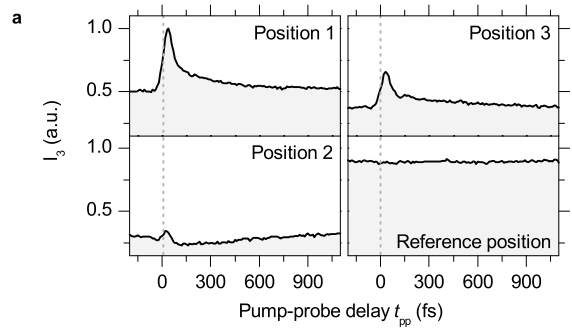
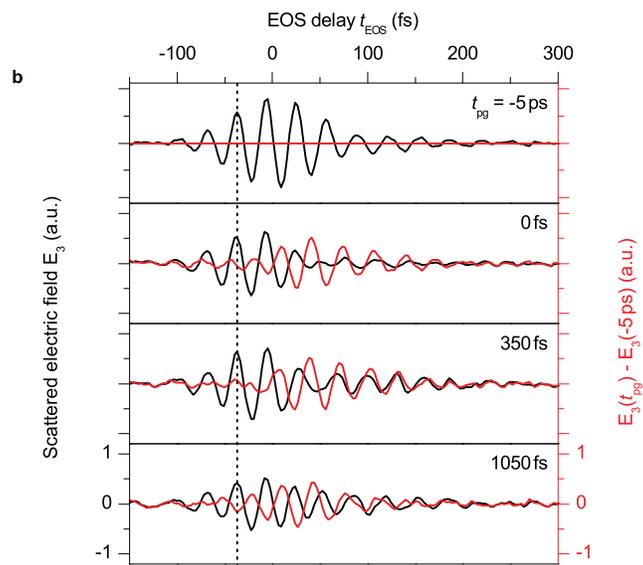
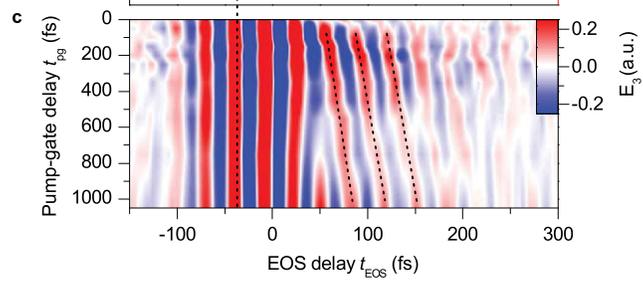

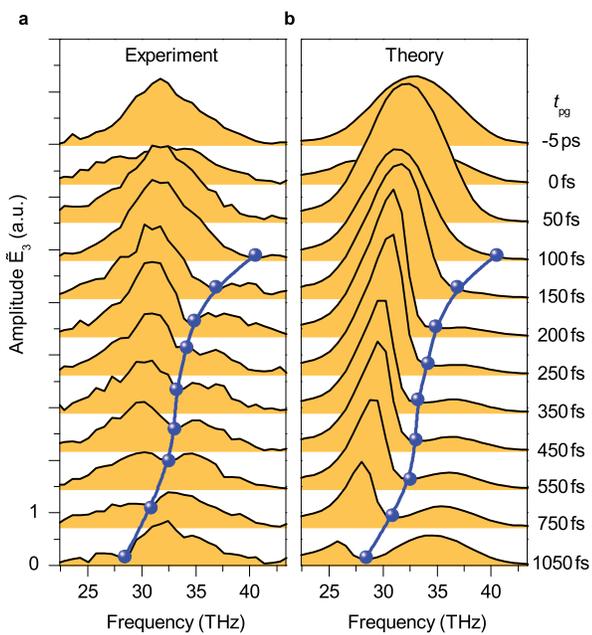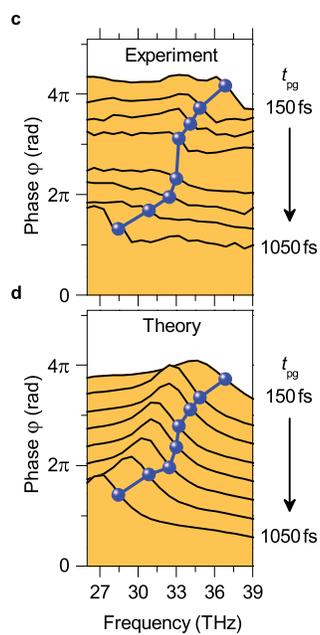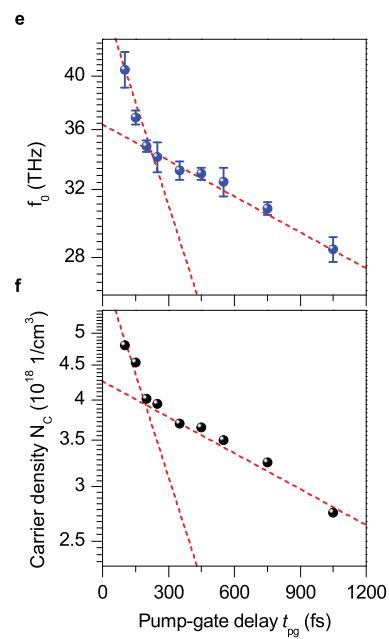

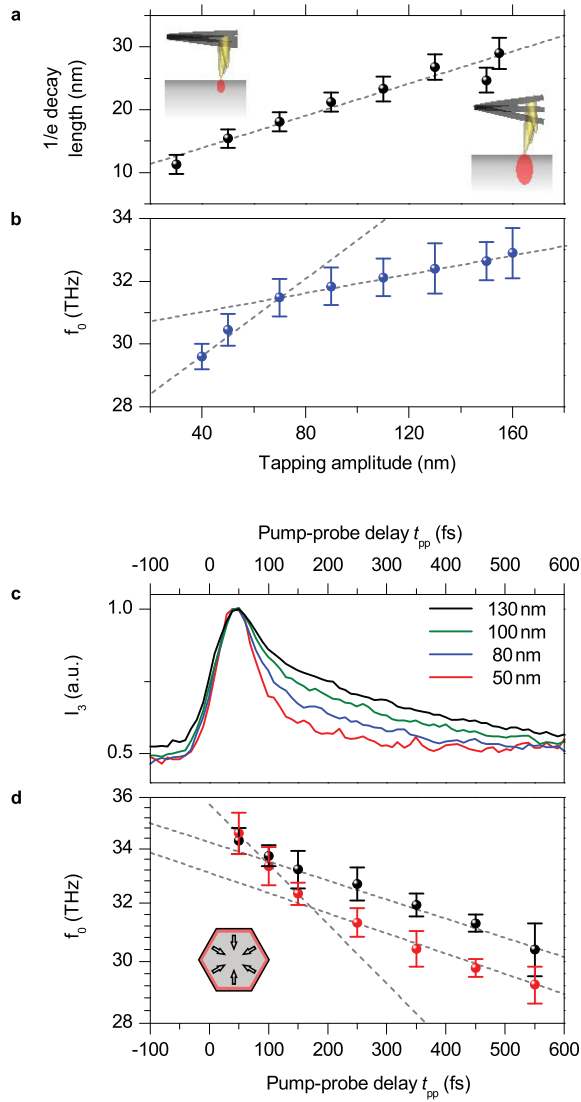